\documentstyle[epsfig,fleqn,amsmath,amssymb]{mn}

\newif\ifAMStwofonts
\AMStwofontstrue     

\ifoldfss
  \ifCUPmtlplainloaded \else
    \NewTextAlphabet{textbfit} {cmbxti10} {}
    \NewTextAlphabet{textbfss} {cmssbx10} {}
    \NewMathAlphabet{mathbfit} {cmbxti10} {} 
    \NewMathAlphabet{mathbfss} {cmssbx10} {} 
  \fi
  \ifAMStwofonts
    \ifCUPmtlplainloaded \else
      \NewSymbolFont{upmath} {eurm10}
      \NewSymbolFont{AMSa} {msam10}
      \NewMathSymbol{\upi}     {0}{upmath}{19}
      \NewMathSymbol{\umu}     {0}{upmath}{16}
      \NewMathSymbol{\upartial}{0}{upmath}{40}
      \NewMathSymbol{\leqslant}{3}{AMSa}{36}
      \NewMathSymbol{\geqslant}{3}{AMSa}{3E}

    \fi
  \fi
\fi 

\ifnfssone
  \newmathalphabet{\mathit}
  \addtoversion{normal}{\mathit}{cmr}{m}{it}
  \addtoversion{bold}{\mathit}{cmr}{bx}{it}
  \newmathalphabet{\mathbfit} 
  \addtoversion{normal}{\mathbfit}{cmr}{bx}{it}
  \addtoversion{bold}{\mathbfit}{cmr}{bx}{it}
  \newmathalphabet{\mathbfss} 
  \addtoversion{normal}{\mathbfss}{cmss}{bx}{n}
  \addtoversion{bold}{\mathbfss}{cmss}{bx}{n}
  \ifAMStwofonts
    \ifCUPmtlplainloaded \else
      \UseAMStwoboldmath
      \makeatletter
      \new@mathgroup\upmath@group
      \define@mathgroup\mv@normal\upmath@group{eur}{m}{n}
      \define@mathgroup\mv@bold\upmath@group{eur}{b}{n}
      \edef\UPM{\hexnumber\upmath@group}
      \new@mathgroup\amsa@group
      \define@mathgroup\mv@normal\amsa@group{msa}{m}{n}
      \define@mathgroup\mv@bold\amsa@group{msa}{m}{n}
      \edef\AMSa{\hexnumber\amsa@group}
      \makeatother
      \mathchardef\upi="0\UPM19
      \mathchardef\umu="0\UPM16
      \mathchardef\upartial="0\UPM40
      \mathchardef\leqslant="3\AMSa36
      \mathchardef\geqslant="3\AMSa3E
    \fi
  \fi
\fi 

\ifnfsstwo
  \DeclareMathAlphabet{\mathbfit}{OT1}{cmr}{bx}{it}
  \SetMathAlphabet\mathbfit{bold}{OT1}{cmr}{bx}{it}
  \DeclareMathAlphabet{\mathbfss}{OT1}{cmss}{bx}{n}
  \SetMathAlphabet\mathbfss{bold}{OT1}{cmss}{bx}{n}
  \ifAMStwofonts
    \ifCUPmtlplainloaded \else
      \DeclareSymbolFont{UPM}{U}{eur}{m}{n}
      \SetSymbolFont{UPM}{bold}{U}{eur}{b}{n}
      \DeclareSymbolFont{AMSa}{U}{msa}{m}{n}
      \DeclareMathSymbol{\upi}{0}{UPM}{"19}
      \DeclareMathSymbol{\umu}{0}{UPM}{"16}
      \DeclareMathSymbol{\upartial}{0}{UPM}{"40}
      \DeclareMathSymbol{\leqslant}{3}{AMSa}{"36}
      \DeclareMathSymbol{\geqslant}{3}{AMSa}{"3E}
    \fi
  \fi
\fi 

\ifCUPmtlplainloaded \else
  \ifAMStwofonts \else 
    \def\upi{\pi}
    \def\umu{\mu}
    \def\upartial{\partial}
  \fi
\fi

\title{Deflection of jets induced by jet-cloud \& jet-galaxy interactions}

\author[S. Mendoza \& M.S. Longair]
       {S. Mendoza \& M.S. Longair\\
        Cavendish Laboratory, Madingley Rd., Cambridge CB3 OHE, U.K.}

\pagerange{\pageref{firstpage}--\pageref{lastpage}}
\pubyear{2000}

\begin{document}

\maketitle

\label{firstpage}


\begin{abstract}
   The model first introduced by Raga \& Cant\'o \shortcite{raga96}
in which astrophysical jets are deflected on passing through an
isothermal high density region is generalised by taking into account
gravitational effects on the motion of the jet as it crosses the high
density cloud.  The problem is also generalised for relativistic jets
in which gravitational effects induced by the cloud are neglected. Two
further cases, classical and relativistic,  are discussed for the cases
in which the jet is deflected on passing through the interstellar gas of a
galaxy in which a dark matter halo dominates the gravitational potential.
The criteria for the stability of jets due to the formation of internal
shocks are discussed.
\end{abstract}



\section{Introduction}

  According to the  standard model for Fanaroff-Riley type II radio
sources,  a pair of relativistic jets of electrons is ejected in opposite
directions  from a very compact region in the nucleus of an elliptical
galaxy \cite{blandford90}.  The jets expand adiabatically through the
interstellar medium of the host galaxy and the surrounding intergalactic
medium.  Collimation of the jets is provided by the presence of a
cavity or ``cocoon'' surrounding the jet, which maintains the pressure
of the intergalactic medium in balance with that of the relativistic gas
within the jet.  In this model, the jets maintain a straight trajectory.
However, observations have shown that jets often bent.

   Dramatic changes in morphology are found in radio trails sources
(for example NGC 1265), in which the jets have a semicircular shape,
with the host galaxy at the pole.  The curvature of the jet is attributed
to the motion of the host galaxy through the intracluster medium which
results in  a significant ram pressure acting on the jet \cite{begelman84}.

  Some deflections may result from a combination of kinetic and
geometrical effects.  For example, the precession of jets about a
defined axis can cause the jet to be curved, although any fluid element
in the jet always follows a straight trajectory.  Another example is the
proper motion of a galaxy through the intergalactic medium produced by
the gravitational influence of a companion galaxy \cite{begelman84}.
Again, each fluid element in the jet follows a straight trajectory.
However, since the galaxy producing the jet moves in an orbit about its
companion, the jet appears to be curved.

  Another way of inducing deflections in radio jets is if the beam passes
through a region of interstellar or intergalactic gas in which there is a
significant pressure gradient.  This might occur if the jets  pass through
the interstellar medium of a large galaxy or diffuse intergalactic cloud.
There are a few cases in which this may have occurred.  In the case of
3C34 Best, Longair \& R\"ottgering \shortcite{best97} found evidence
that the jet had interacted strongly with the interstellar medium of a
galaxy which happened to lie in the path of the jet.  More evidence for
misalignments between one of the hotspots of a radio galaxy and the host
galaxy was found in 3C324 and 3C368 \cite{best98}.  In these objects,
no galaxy has been found in the path of the jets, but there is strong
evidence for the presence of high density regions from the optical and
infrared observations. For sources in which the jet has given rise to
optically aligned emission regions, the optical emission is most probably
produced by shock emission \cite{best00}. The present study was motivated
by these observations, the objective being to find a simple prescription
for how significant the deflections of the radio jets could be if they
passed through the interstellar medium of an intervening galaxy or gas
cloud.  In addition, we have analysed the stability of these deflections
against to the formation of internal shocks.  This stability analysis
will be published elsewhere, but we discuss the results at the end of
this article.

  The work presented here is an extension of the model first proposed by
Cant\'o \& Raga \shortcite{canto96} and Raga \& Cant\'o \shortcite{raga96}
in order to describe how deflections of radio jets are produced in
jet/cloud interactions.  The steady interaction of a jet with an
isothermal sphere is analysed in two ways.  Firstly we describe the
steady interaction of a classical jet (i.e. a jet for which its internal
velocities and thermal effects are not relativistic) with an isothermal
cloud in which the self gravity of the cloud is taken into account.
This is similar to the analysis of Raga \& Cant\'o \shortcite{raga96},
but generalised in order to take into account  the effects of gravity.
We also analyse the steady interaction of a relativistic jet with an
isothermal sphere in which the gravity of the cloud is not taken into
account.  In this case, it possible to integrate the relevant equations
exactly.  Secondly, we consider a more relevant case for galaxies, in
which it is assumed that all the gas in the galaxy is in hydrostatic
equilibrium within a dark matter halo.  Again, the cases of classical
and relativistic jets are discussed.

\section{Jet/Cloud interactions}

  The interaction of an astrophysical jet with a cloud, in which the
characteristic size of the cloud is much greater than the jet radius, has
been studied in its initial stages by Raga \& Cant\'o \shortcite{raga95}.
They showed that initially the jet slowly begins to bore a passage into
the cloud.  Eventually a stationary situation is achieved in which the
jet  penetrates the cloud and escapes from it in a direction which is
different from the original jet trajectory.

  Once the steady state is reached, the trajectory is determined by
maintaining pressure equilibrium with the surrounding environment.
In other words, as the material in the jet moves, it adjusts its pressure
in such a way that it is in equilibrium with the internal pressure of
the cloud.  Since the expansion of the jet is adiabatic, and a steady
state has been reached, Bernoulli's equation can be used to describe the
trajectory of the jet as it passes through the cloud. In what follows the
interaction of a jet with a cloud of characteristic radius much greater
than the diameter of an adiabatic jet is analysed, once the steady state
configuration has been reached.

\section{Classical Analysis}

  In the classical case, the motion of a fluid element in the
jet is determined by Euler's equation:

\begin{equation}
   \frac{\partial {\boldsymbol v}}{\partial t} + \boldsymbol{v}\! \cdot
        \!{ \mathbf{grad} } \: \boldsymbol{v} = -\frac{1}{\rho} {
        \mathbf{grad} } \: p - { \mathbf{grad}} \: \phi,
\label{eq.3.1}
\end{equation}

\noindent where $\boldsymbol{v}$ is the velocity of the flow, and $p$
and $\rho$ are pressure and density.  The gravitational force produced
by the cloud  is described in the term which contains the gravitational
potential $\phi$.  The left hand side of eq.(\ref{eq.3.1}) is the
force per unit mass experienced by a fluid element as it moves.
Since all quantities depend only  on the distance from the cloud's
centre $\boldsymbol{r}$, vector multiplication of the radius vector
$\boldsymbol{r}$ with eq.(\ref{eq.3.1}) implies $\boldsymbol{r} \! \times
\! {\rm d}\boldsymbol{v}/{\rm d}t \! = \! 0$. In other words, the specific
angular momentum $\boldsymbol{l} \! = \! \boldsymbol{r} \!  \times \!
\boldsymbol{v} $ is conserved as the fluid moves.  Since the radius
vector is perpendicular to the angular momentum vector, the motion is
two dimensional, and so, polar coordinates ($r$,$\varphi$) are used in
the following analysis.

	Consider a situation in which the jet enters the cloud
parallel to the $x$ axis at a distance $r_0$, so that its velocity
vector is initially given by:

\begin{equation}
   \boldsymbol{v}_0 = -v_0\boldsymbol{e}_x = -v_0\left(\cos\varphi_0
        \, \boldsymbol{e}_r - \sin\varphi_0 \, \boldsymbol{e}_\varphi
        \right),
\label{eq.3.2}
\end{equation}

\noindent where $\boldsymbol{e}_x$, $\boldsymbol{e}_r$ and
$\boldsymbol{e}_\varphi$ are unit vectors in the directions $x$,
$r$ and $\varphi$ respectively.  Because angular momentum is conserved
and the motion is two dimensional, the velocity is most simply written
using eq.(\ref{eq.3.2}) as:

\begin{equation}
   \boldsymbol{v} = v_r\boldsymbol{e}_r + \frac{r_0}{r}v_0\sin\varphi_0
                    \,\boldsymbol{e}_\varphi, 
\label{eq.3.3}
\end{equation}
  
\noindent in which $v_r\!=\!{\rm d}r/{\rm d}t$ represents the velocity in the
radial direction.  

   Since the steady flow of inside the jet expands adiabatically, we can
calculate its path of by means of Bernoulli's equation:

\begin{equation}
  \frac{1}{2} \int{{\rm d}v^2} +\int{{\rm d}w} + \int{{\rm d}\phi} = 0,
\label{eq.3.4}
\end{equation}

\noindent in which the line integrals are taken from the initial position
of a given fluid element to its final one.  The enthalpy per unit mass
$w$ of the flow in the jet is given by $w\! =\! \Gamma^{-1} p/\rho$
and $\Gamma \! \equiv \!  (\gamma-1)/\gamma$ for a gas with polytropic
index $\gamma$.

  Substituting eq.(\ref{eq.3.3}) and eq.(\ref{eq.3.2}) into the first
integral of eq.(\ref{eq.3.4}) gives:

\begin{equation}
  \int{{\rm d}v^2} = \left\{\frac{r_0}{r} \frac{v_0\, \sin\varphi_0}{r}
	       \frac{{\rm d}r}{{\rm d}\varphi} \right\}^2 + v_0^2\left\{
	       \left(\frac{r_0}{r}\right)^2 \sin^2 \negthinspace \varphi_0
	       -1 \right\},
\label{eq.3.5}
\end{equation}

\noindent where we have used the fact that along the jet trajectory
${\rm d}r/v_r \! = \! r{\rm d}\varphi / v_\varphi$.  Since the gas in
the jet obeys a polytropic equation of state, we obtain for the second
integral in eq.(\ref{eq.3.4}):

\begin{equation}
   \int{{\rm d}w} = \frac{c_0^2}{\gamma \Gamma} \left\{
   	     \left(\frac{p}{p_0}\right)^\Gamma - 1 \right\}, 
\label{eq.3.6}
\end{equation}

\noindent in which $c_0^2$ is the initial sound speed of the jet material.
The integral for the gravitational potential produced by the
self-gravitating cloud is obtained from:

\begin{equation}
   \int{{\rm d}\phi} = G\int{\frac{M(r)}{r^2} \, {\rm d}r} = 4\pi G \int{
	  \frac{{\rm d}r}{r^2} \int^r_0{\xi^2 \rho_{c}(\xi) \, {\rm
	  d}\xi} },
\label{eq.3.7}
\end{equation}

\noindent where $M(r)$ and $\rho_c(r)$ represent the mass
and density of the cloud at a distance $r$.  Substitution of
eqs.(\ref{eq.3.5})-(\ref{eq.3.7}) into eq.(\ref{eq.3.4}) gives the
equation for the path of the jet as it expands:

\begin{equation}
 \begin{split}
   \frac{{\rm d}\eta}{{\rm d}\varphi} & = \pm \frac{1}{\sin\varphi_0} 
      \Bigg\{ 1 - \eta^2 \sin^2\negthickspace\varphi_0 - 
      \frac{2}{ \gamma \Gamma M^2_0} \times \\
      &  \times \left[ \left( \frac{p}{p_0} \right)^\Gamma - 1 \right]
      - \frac{8\pi G}{M^2_0 c_0^2}  \int{ \frac{{\rm d}r}{r^2} 
      \int^r_0{\xi^2 \rho(\xi) \, {\rm d}\xi} }\Bigg\}^{1/2}_, \\
\end{split}
\label{eq.3.8}
\end{equation}

\noindent in which $\eta \! = r_0 / r$ and $M_0$ is the initial Mach
number of the flow in the jet.   The positive and negative choices
for the value of the derivative $ {\rm d}\eta/{\rm d}\varphi $ in
eq.(\ref{eq.3.8}) have to be chosen with care.  For example, if we
consider the case in which no gravity and no pressure gradients are
taken into account (i.e. the last two terms on the right hand side of
eq.(\ref{eq.3.8}) are zero, which corresponds to a straight trajectory)
the derivative $ {\rm d}\eta/{\rm d}\varphi \! < \! 0 $ for $\eta_*
\! < \!  1 / \sin \varphi_0$ and vice versa.  The equality $\eta_* \! = \!
1/ \sin\varphi_0$ corresponds to the point at which a given fluid element
in the jet reaches the $y$ axis during its motion in this particular case.

  In the limit of high initial supersonic motion ($M_0 \!  \gg \!  1$)
the third and fourth terms on the right hand side of eq.(\ref{eq.3.8})
are important only when $\eta \! = \! 1/\sin\varphi_0 $ and  we can
simplify eq.(\ref{eq.3.8}) by making an expansion about this point.
In general terms, if:

\begin{gather}
   \left( \frac{p}{p_0} \right)^\Gamma  = \alpha + \beta \eta 
      \sin\varphi_0 + \zeta \eta^2 \sin^2\negthickspace\varphi_0, 
                                                  \label{eq.3.9} \\
   \intertext{and}
   \phi - \phi_0 = 4 \pi G \left( \tilde\alpha + \tilde\beta \eta 
      \sin\varphi_0 + \tilde\zeta \eta^2 \sin^2 \negthickspace
      \varphi_0 \right)
        					   \label{eq.3.10} 
\end{gather}

\noindent are expansions of the pressure and gravitational potential
respectively around $\eta \! = \! 1/\sin\varphi_0$, then eq.(\ref{eq.3.8})
takes the form:

\begin{gather}
   \frac{{\rm d}\eta}{{\rm d}\varphi} = \pm \frac{1}{\sin\varphi_0}
      \left\{ a + b \eta + e \eta^2 \right\}^{1/2}_,
      						\label{eq.3.11} \\
  \intertext{in which:}
   a \equiv 1 - \frac{2 \left( \alpha -1 \right)}{ \gamma 
     \Gamma M_0^2 } - \frac{8\pi G}{M_0^2 c_0^2} 
     \tilde\alpha, 
          					\notag \\
   b \equiv - \left( \frac{ 2 \beta }{ \gamma \Gamma M_0^2 } +
      \frac{8 \pi G}{M_0^2 c_0^2} \tilde\beta \right) \sin \varphi_0,
   						\notag \\
   e \equiv -\left( 1 + \frac{ 2 \zeta }{ \gamma \Gamma M_0^2 } +
      \frac{ 8 \pi G}{M_0^2 c_0^2} \tilde\zeta \right) \sin^2
      \negthinspace \varphi_0,
   						\notag 
\end{gather}

\noindent to second order in  $\eta \sin \varphi_0$.  For the cases
considered in this analysis, the general solution of eq.(\ref{eq.3.11})
is:

\begin{equation}
 \begin{split}
   \eta=\frac{1}{2e} & \left\{\sqrt{\Delta}\sin\left[\sqrt{-e}\left(
       \frac{\varphi_0 -\varphi}{\sin\varphi_0} \ +
       \right. \right. \right. \\
   & +  \left.\left.\left. \frac{1}{\sqrt{-e}} \arcsin
      \frac{2e+b}{\sqrt{\Delta}} \right) \right] -b \right\}
\end{split}
\label{eq.3.12}
\end{equation}

\noindent with $\Delta \! \equiv \! b^2 -4ae$. The angle between the
velocity vector of the jet streamline and the $x$ coordinate axis on
its way out of the cloud (deflection angle) can be calculated from the
relation $\tan\psi \! = \!  (v_y/v_x)_{\rm exit}$, in other words,

\begin{equation}
   \psi=\arctan\left(\frac{\sin\varphi_{\rm e}\left({\rm d}\eta/{\rm d}
	\varphi\right)_{\rm e}	- \cos\varphi_{\rm e} } { \cos\varphi_{\rm
	e} \left({\rm d}\eta / {\rm d}\varphi \right)_{\rm e} +
	\sin\varphi_{\rm e} }\right) + \pi,
\label{eq.3.13}
\end{equation}

\noindent where the exit azimuthal angle $\varphi_{\rm e}$ is given by:

\begin{equation}
   \varphi_e = \varphi_0 + \frac{ \sin\varphi_0 }{ \sqrt{-e} } \left\{ 2
      \arcsin \frac{ 2e + b }{ \sqrt{ \Delta } }  + \pi \right\},
\label{eq.3.14}
\end{equation}

\noindent provided the deflections are not large. The derivative $( {\rm
d}\eta / {\rm d}\varphi )_{\rm e}$ is evaluated at $\eta \! = \! 1$,
with a negative choice of sign in eq.(\ref{eq.3.8}).

\subsection{Isothermal Cloud}
\label{isoclass}

  Let us consider the case of an isothermal cloud, for which the density
in the cloud $\rho_{\rm cloud}=\xi /r^2$.  In other words, because the
jet and cloud are maintained in pressure balance, the pressure on the
jet is given by:

\begin{equation}
   \frac{p}{p_0} = \left( \frac{r_0}{r} \right)^2.
\label{eq.3.15}
\end{equation}

\noindent   For this isothermal case, it is easy to verify that:

\begin{alignat}{2}
   \alpha & = \frac{ 1 - \Gamma \left( 3 - 2\Gamma \right) }{ \sin^{2\Gamma}
      \negthickspace \varphi_0}, & \qquad  \tilde\alpha & = \xi
      \ln(\sin\varphi_0) + \frac{3}{2}\xi,
						\notag \\
   \beta & = \frac{ 4 \Gamma \left( 1 - \Gamma \right) }{ \sin^{ 2 \Gamma }
      \negthinspace \varphi_0 }, & \qquad \tilde\beta & = 2 \xi,
      						\label{eq.3.16} \\
   \zeta & = \frac{ \Gamma \left( 2\Gamma - 1 \right) }{ \sin^{ 2 \Gamma }
      \negthinspace \varphi_0 }, & \qquad \tilde\zeta & = \frac{ \xi }{ 2 }.
						\notag \\
\end{alignat}

  This solution corresponds to that found by Raga \& Cant\'o
\shortcite{raga96} for the case in which no gravity is present,
i.e. $\tilde\alpha \! = \! \tilde\beta \! = \! \tilde\zeta \! =
\! 0$.  From the solutions obtained above in eq.(\ref{eq.3.11}) and
eq.(\ref{eq.3.16}) it follows that the dimensionless quantity \( \Lambda
\! \equiv \! G \xi / M_0^2 c_0^2 \) is a number that  parametrises the
required solution.

  The deflection of jets in isothermal clouds may be important for
interstellar molecular clouds and the jets associated with Herbig-Haro
objects. For this case we can obtain a value for the parameter $\Lambda$.
If we adopt  a particle number density of $n_{\rm H} \! \sim \! 10^{2} \,
{\rm cm}^{-3}$, and a temperature $T \! \sim 10 \, {\rm K}$ for a cloud
with radius $r_0 \! \sim 1 \,  {\rm pc} $ \cite{spitzer98,hartmann98},
then

\begin{equation}
  \Lambda  \sim  \frac{ 10^ {-2} }{ M_0^2 } \left( \frac{ r_0 }{ 1 \,
    \mathrm{pc} } \right)^2 \left( \frac{ n_0 }{ 10^2 \, \mathrm{cm}^{-3} }
    \right) \left( \frac{ T }{ 10 \, \mathrm{K} } \right)^{-1}.
\label{eq.3.16b}
\end{equation}

\noindent The same calculation can be made for the cases of radio
jets interacting with the gas inside a cluster of galaxies.  For this
case, typical values are \( n_{\rm H} \sim 10^{-2}\, {\rm cm}^{-3},
\ T \sim 10^7\, {\rm K} \text{ and } r_0 \sim 100 \, {\rm kpc} \)
\cite{longair92,longair98}.  With these values, the parameter  \(
\Lambda \sim 10^{-2} / M_0^2 \), exactly as eq.(\ref{eq.3.16b}).
The parameter \( \Lambda \) is an important number which can be obtained
by dimensional analysis.  For, the problem in question is characterised
by the gravitational constant \( G \), a ``characteristic length'' \(
r_0 \) and the values of the velocity of the jet and the density \( v_0
\) and \( \rho_0 \) respectively.  Three independent dimensions (length,
time and mass) describe the whole hydrodynamical problem.  Since four
independent physical quantities (\( G,\, \rho_0,\, v_0,\,\text{and}\,
r_0 \)) are fundamental for the problem we are interested, the \mbox{Buckingham \( \Pi \)--Theorem} of dimensional analysis requires the
existence of only one dimensionless parameter \( \Lambda = G \xi /
M_0^2 c_0^2 \), exactly as in the analytic solution. The fact that jets
are formed in various environments such as giant molecular clouds and
the gaseous haloes of clusters of galaxies with the same values of the
dimensionless parameter \( \Lambda \) provides a clue as to why the jets
look the same in such widely different environments.  

  From its definition, the parameter \( \Lambda \) can be rewritten as \(
\Lambda = ( 3 / 4 \pi ) ( G { \mathsf{M} } / r ) ( 1 / v_0^2 ) \), where
\( { \mathsf{M} } \) is the mass within a sphere of radius \( r_0 \).
This quantity is roughly the ratio of the gravitational potential energy
from the cloud acting on a fluid element of the jet, to its kinetic energy
at the initial position \( r_0 \). The parameter \( \Lambda \) is thus
an indicator of how large the deflections due to gravity are going to
affect the trajectory of the jet. The bigger the number \( \Lambda \),
the more important the deflection caused by gravity will be.  In other
words, when the parameter \( \Lambda \gg 1 \) the jet becomes ballistic and
bends towards the centre of the cloud.  When \( \Lambda \ll 1 \) the
deflections are dominated by the pressure gradients in the cloud and the
jets bend away from the centre of the cloud.

\begin{figure*}
   \begin{center}
      \epsfig{file=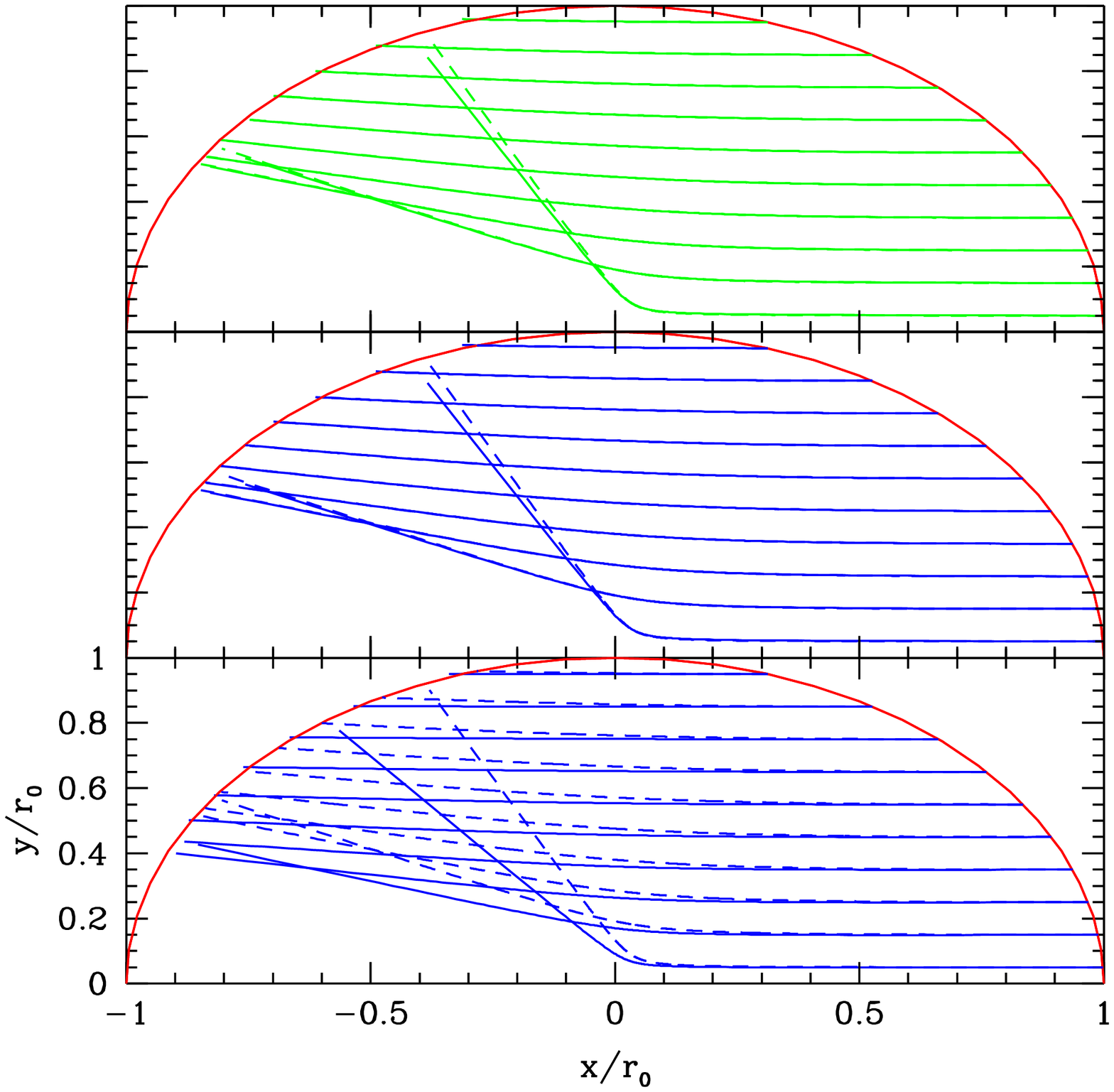,height=8.45cm}
      \epsfig{file=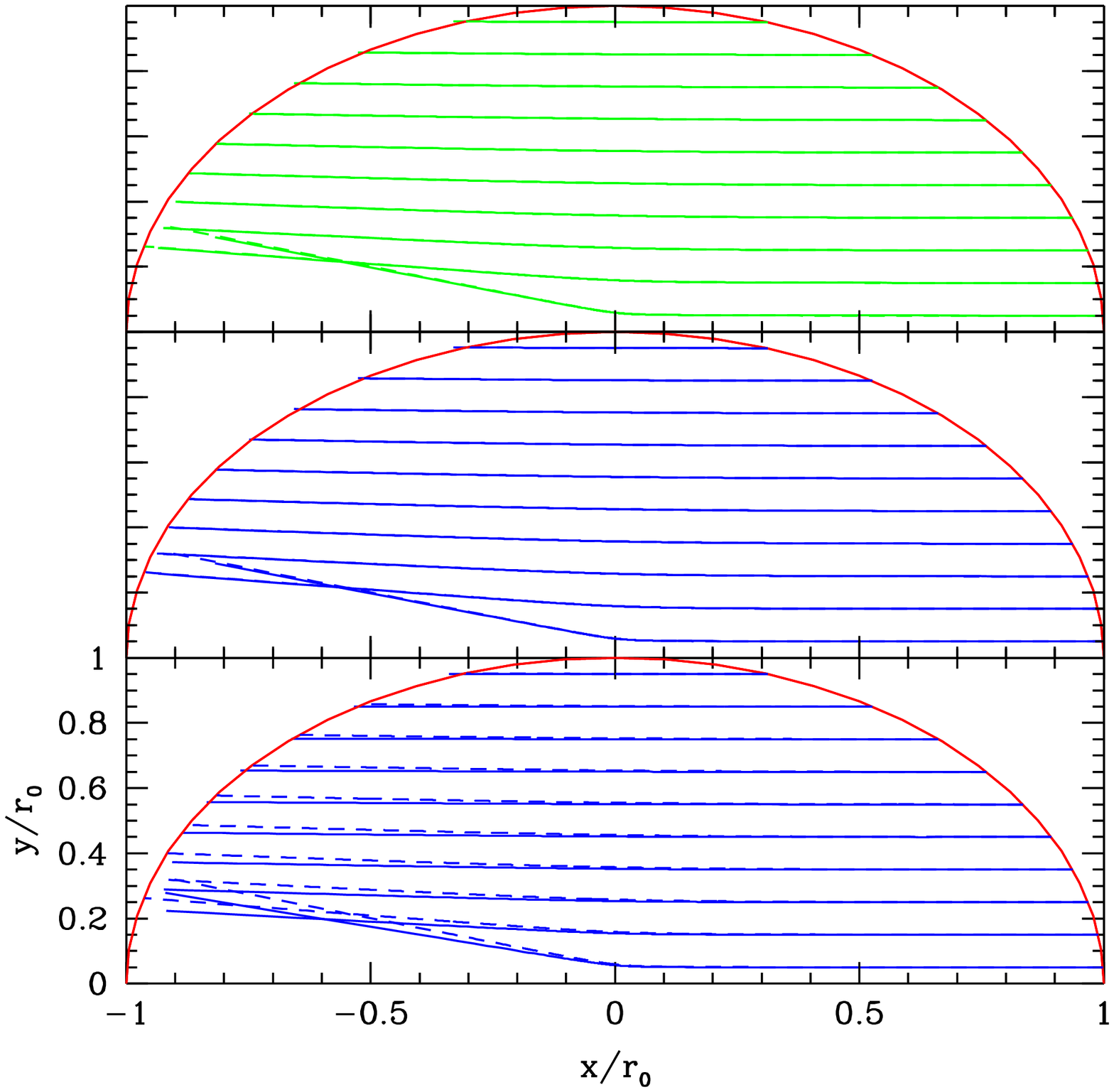,height=8.45cm}
   \end{center}
    \caption{Deflection produced in a jet due to the interaction with
	     an isothermal cloud (semicircle) of radius $r_0$. The
	     jet penetrates the cloud from the right, parallel to
	     the $x$ axis.  Different trajectories are shown in each
	     diagram for different initial heights of $y/r_0\!=\!0.05$,
	     $0.15$,...,$0.95$ as measured from the $x/r_0$ axis.
	     In each figure the top diagram corresponds to the
	     case in which gravitational effects are not considered
	     {\protect\cite{raga96}}.  The central and bottom diagrams
	     represent trajectories for which gravitational effects are
	     taken into account and the parameter $\Lambda$ has values
	     of $10^{-6}$, and $0.01$ respectively  in units of the
	     square of the  initial Mach number $M_0^{2}$ of the jet
	     (see text).  A polytropic index $\gamma \! = \! 5/3$ for
	     the flow in the jet was assumed.  The diagrams at the left
	     and right correspond to initial Mach numbers for the jet
	     flow of $M_0 \!  = \! 5$ and $M_0 \! = \! 10$ respectively.
	     The dashed lines in the graphs represent the direct numerical
	     integration of the equation of motion.  The continuous lines
	     are the analytic approximations discussed in this article.}
\label{fig.3.1}
\end{figure*}

  Fig.(\ref{fig.3.1}) shows plots for three different values of $\Lambda$
with initial Mach numbers of $M_0 \! = \!  5$ and $M_0 \! = \! 10$.
A comparison with a numerical integration of eq.(\ref{eq.3.11}) using
a fourth--order Runge--Kuta method is also presented in the figures by
dashed lines.  This comparison shows that as long as the deflections
are sufficiently small, or as long as the Mach number of the flow in the
jet is sufficiently large, the analytic approximations discussed above
are a good approximation to the exact solution.

\subsection{Gas on Dark Matter Halo}
\label{classdark}

    Let us consider next the case of a galaxy dominated by a dark matter
halo for which the mass density is given by the relation \cite{binney87}:

\begin{equation}
   \rho_d = \frac{ \rho_{d_\star} }{ 1 + \left( \frac{r}{a} \right)^2 },
\label{eq.3.18}
\end{equation}

\noindent in which $a$ is the core radius and quantities with a star refer
to the value at the centre of the galaxy. 

  The potential resulting from such a density profile can be calculated
by means of eq.(\ref{eq.3.7}),

\begin{multline}
   \phi_d -\phi_{d_\star} = 4 \pi G \rho_{d_\star} a^2 \times \\
      \times \left\{ \frac{1}{2} 
      \ln \left[ 1 + \left( \frac{r}{a} \right)^2 \right] + \frac{a}{r} 
      \arctan \left( \frac{r}{a} \right) -1 \right\},
\label{eq.3.19}
\end{multline}

\noindent in which the value of the gravitational potential
$\phi_{d_\star}$ has been evaluated at the centre of the galaxy $r_\star
\! = \! 0$.  If the gas in the galaxy is in hydrostatic equilibrium
with the dark matter halo, then $ { \mathbf{grad} }\: p \! = -\rho \,
{ \mathbf{grad} }  \: \phi_d $.  In this case, the enthalpy of the
isothermal gas is given by:

\begin{gather}
   w - w_\star = -\phi_d + \phi_{d_\star} = c^2_\star \ln \left(
      \frac{p}{p_\star} \right),
						\label{eq.3.20} \\
   \intertext{and so the pressure is:}
   p = p_\star \, \exp \left\{ \left( -\phi_d + \phi_{d_\star} \right) /
      c^2_0 \right\}.
					         \label{eq.3.21} 
\end{gather}

\begin{figure*}
   \begin{center}
      \epsfig{file=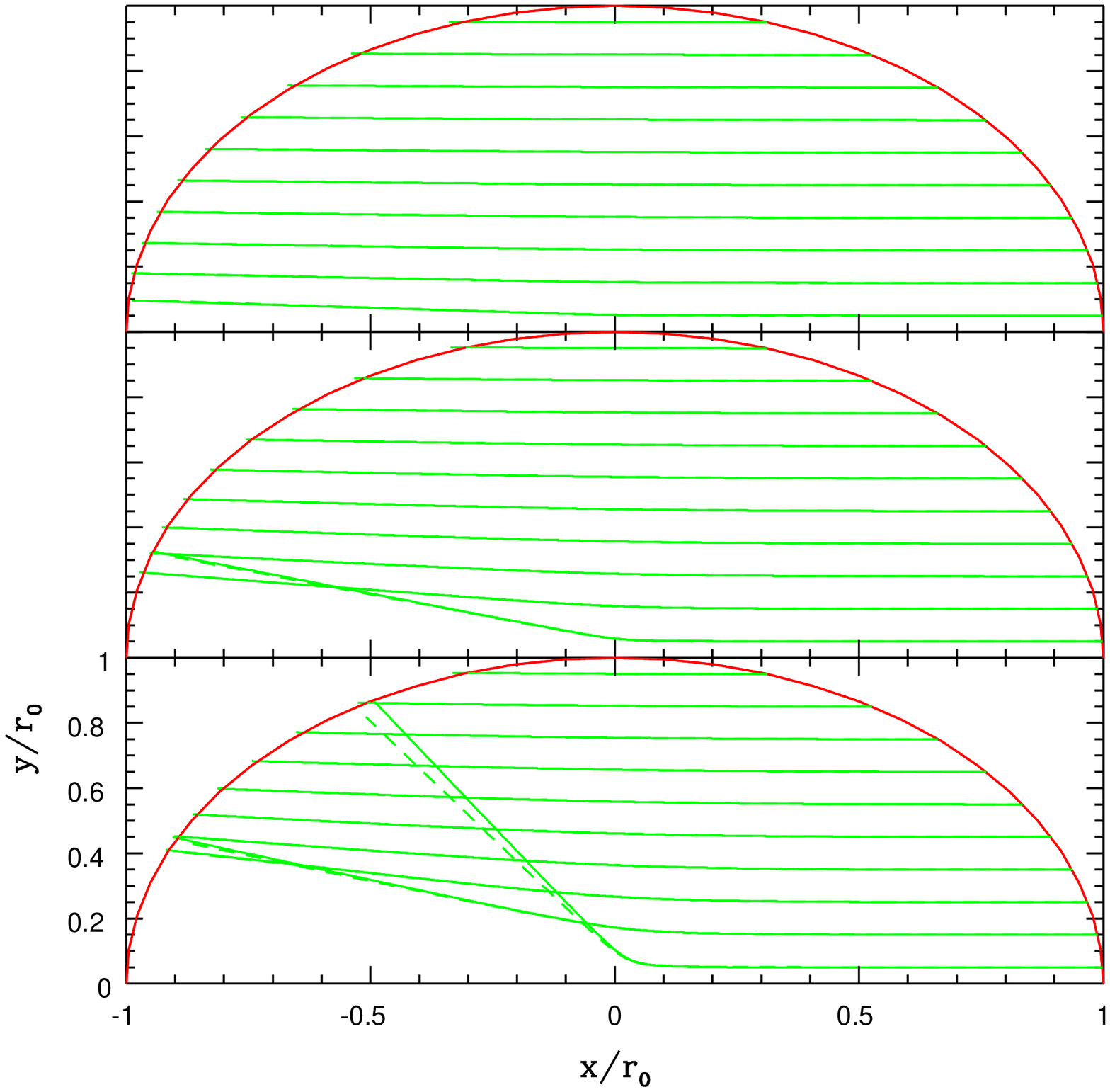,height=8.45cm}
      \epsfig{file=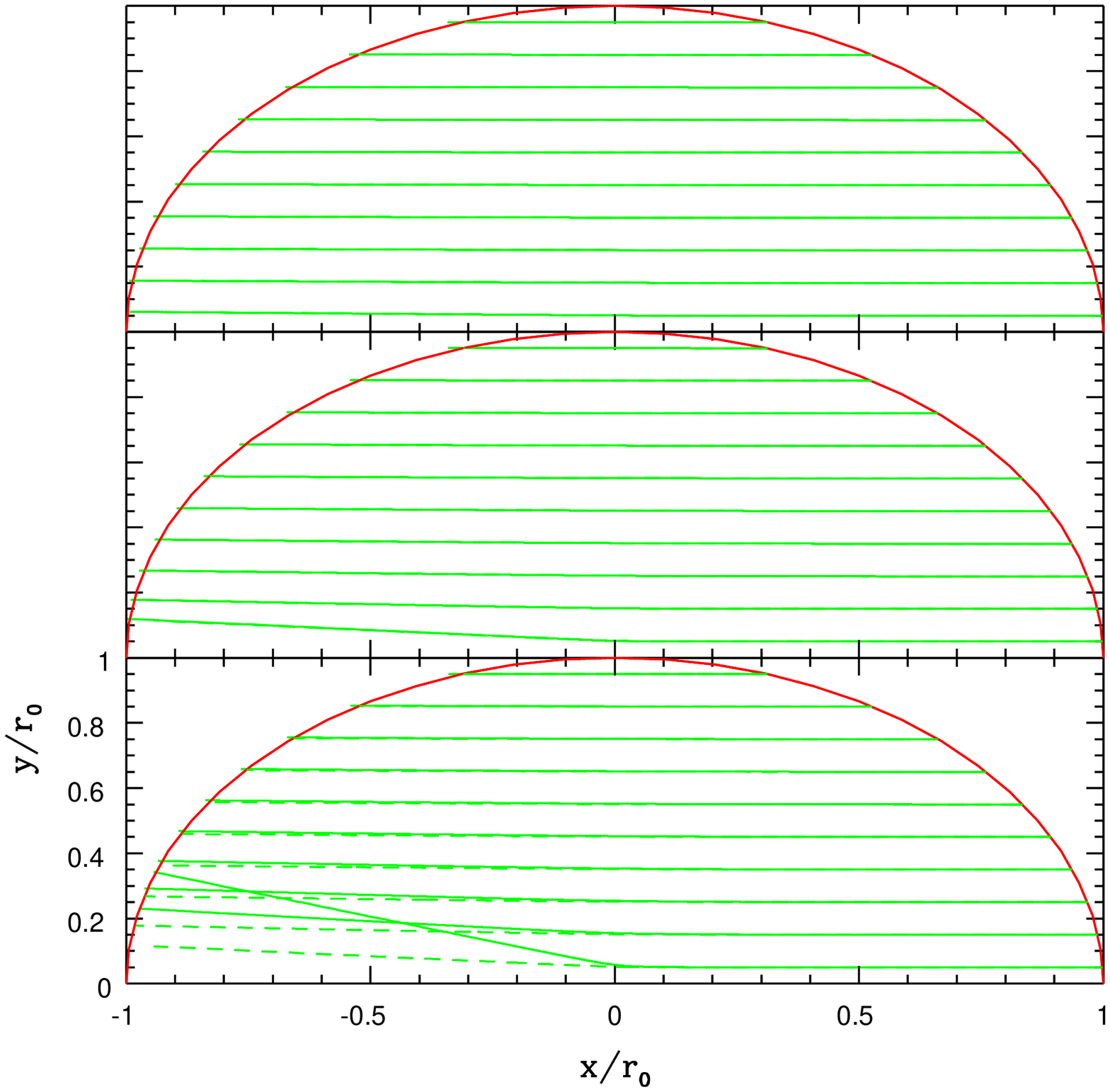,height=8.45cm}
   \end{center}
    \caption{Deflection produced in a jet as it travels across a galaxy,
             for which its gravitational potential is dominated by a dark
             matter halo.  The jet penetrates the galaxy parallel to
             the $x/r_0$ axis.  Various trajectories are shown in each
             diagram for different initial heights $y/r_0 \! = \! 0.5$,
             $0.15$,...,$0.95$ measured from the $x/r_0$ axis.  The left
             and right diagrams were calculated for the case of a non
             relativistic jet with an initial  Mach number of $M_0 \!=
             \! 10$ and $M_0 \! = \! 20$ respectively.  For each of
             this diagrams a value of $k \! = \! -1$, $-2$, $-3$ was
             used for the top, middle and bottom panels respectively
             (see text).  The dashed lines in the figures represent
             the direct numerical integration of eq.(\ref{eq.3.11})
             with the pressure given by eq.(\ref{eq.3.21}) for the case
             in which the ratio of the core radius $a$ to the initial
             radius $r_0$ is given by $a/r_0 \! = \! 10^{-3} \ll 1$.
             The continuous lines are analytic approximations found
             under this conditions.}
\label{fig.3.2}
\end{figure*}

   It is possible to simplify the above expressions using the fact
that for astronomical cases $r_0 \! \gg \! a$.  In other words:

\begin{equation}
   \frac{p}{p_0} = \eta^{-k},
\label{eq.3.22}
\end{equation}

\noindent where $ k = - 4 \pi G \rho_{d_\star} a^2 / c_\star^2 $.
In this approximation, the required analytic solutions can be found:

\begin{gather}
   \alpha = \frac{1}{2} \left( \Gamma k + 1 \right) \left( \Gamma k + 2
      \right) \sin^{ \Gamma k } \! \varphi_0,
      							\notag \\
   \beta  = - \Gamma k \left( 2 +  \Gamma k \right) \sin^{ \Gamma k}
      \! \varphi_0, 
      							\label{eq.3.23} \\
   \zeta  = \frac{1}{2}  \Gamma k \left( \Gamma k + 1 \right) 
      \sin^{ \Gamma k} \! \varphi_0, 
      							\notag 
\end{gather}

\noindent where for simplicity it was assumed that $\tilde\alpha
\negmedspace = \negmedspace \tilde\beta \negmedspace = \negmedspace
\tilde\zeta \negmedspace = \negmedspace 0$, in other words we neglect the
gravitational field induced by the mass of the cloud.  Typical
values \cite{binney87} for galaxies are $\rho_\star \! \sim \!
0.1 \,{\rm M}_\odot \, {\rm pc}^{-3} $, $ a \! \sim \! 1 \, {\rm kpc}
$. Taking central values for the gas in the galaxy as $n_\star \! \sim
\! 1 \, {\rm cm}^{-3}$ and $T_\star \! \sim \! 10^5 \, {\rm K} $ then
$k \! \sim \! -10$.  Fig.(\ref{fig.3.2}) presents plots for different
values of $M_0$ and $k$.

  The dimensionless number \( k \) which parametrises the required
solution can be obtained by dimensional analysis (apart from an
unimportant factor of \( - 4 \pi \)) in the same way as the parameter \(
\Lambda \) was obtained in Section \ref{isoclass}.  In this case the
important parameters in the problem are the gravitational constant
\( G \), the characteristic length \( a \) and the sound speed \(
c_\star \).  The number \( k \) can be rewritten as \( k = - \left( 4 /
( 4 - \pi ) \right)  ( G { \mathsf{M} } / a ) ( 1 / c_\star^2 ) \),
where \( { \mathsf{M} } \) is the mass of a sphere with radius \( a \).
This quantity is proportional to the gravitational energy of the cloud
evaluated at the core radius divided by the sonic kinetic energy that
a fluid element in the jet has.  In other words, in the same way as in
Section \ref{isoclass}, the dimensionless number \( k \) is an indicator
as to how big deflections produced by gravity are.

\section{Relativistic Analysis}

  Let us consider the case in which relativistic effects are included
in the  interaction of a relativistic jet and a stratified high density
region. In order to simplify the problem, the self gravity of the cloud
acting on the jet is ignored.  For this case, the relativistic 4-Euler's
equation is:

\begin{equation}
   wu^k\frac{\partial u_i}{\partial x^k} = \frac{\partial p}{\partial
        x^i} - u_i u^k \frac{\partial p}{\partial x^k}
\label{eq.4.1}
\end{equation}

\noindent in which $w$, and $p$ are the enthalpy per unit volume and
the pressure of a given fluid  element in its proper frame of reference
respectively \cite{daufm}.  The 4-velocity $u^k \! = \!  dx^k/ds$ (latin
indices run as 0, 1, 2, 3) is given by ($\gamma,\gamma\boldsymbol{v} /c$),
$c$ is the speed of light and $\gamma$ is the Lorenz factor.
The 4-radius vector $x^k \!= \! (ct,\boldsymbol{r})$ and the metric tensor
$\eta_{kl} \! = \! {\rm diag}(1, -1, -1, -1)$.  The space component of
eq.(\ref{eq.4.1}) can be written as:

\begin{equation}
   \frac{w\gamma^2}{c^2} \left(\frac{\partial \boldsymbol{v}
      }{\partial t} + \boldsymbol{v} \! \cdot \!  { \mathbf{grad} }
      \, \boldsymbol{v} \right) = - { \mathbf{grad} } \, p
      -\frac{\boldsymbol{v}}{c^2} \frac{\partial  p}{\partial  t},
\label{eq.4.2}
\end{equation}

\noindent which is the generalisation of Euler's equation.  As before,
the term in brackets on the left-hand side represents the classical
force per unit mass acting on an element of fluid as it moves.
By considering steady flow and noting that all quantities depend only
on the radial coordinate $r$, vector multiplication of the radius
vector $\boldsymbol{r}$ with eq.(\ref{eq.4.2}) shows that the quantity
$\boldsymbol{l} \! = \! \boldsymbol{r} \times \boldsymbol{v}$ is conserved
as the fluid moves.  This quantity corresponds to the classical equivalent
of the specific angular momentum, but it is not the relativistic angular
momentum which is given by $\boldsymbol{r} \times \gamma\boldsymbol{v}$.
The constancy of $\boldsymbol{l}$ shows that the motion is two dimensional
and so eqs.(\ref{eq.3.2})-(\ref{eq.3.3}) are valid in the relativistic
case as well.

  The trajectory of the jet is described by Bernoulli's equation, which
is given by the time component of eq.(\ref{eq.4.1}) for steady adiabatic
flow by \cite{daufm}:

\begin{equation}
   \int{{\rm d}\left(\frac{\gamma w}{n}\right)} =0.
\label{eq.4.3}
\end{equation}

  The line integral is taken from the initial position of a given fluid
element to its final one.  The particle number per unit proper volume
is $n$ and we assume that the electrons in the jet are ultrarelativistic,
so that the equation of state is given by $p \! = \! e/3$ with $e$ being
the internal energy density of the fluid.  The requirement that the
pressure of the jet equals that of the cloud, together with the fact that
$p \! \varpropto \! n^{4/3} $ gives:

\begin{equation}
  \frac{{\rm d}\eta}{{\rm d}\varphi} = \pm \frac{c}{v_0 \sin \varphi_0}
     \left\{ 1 - \frac{v_0^2}{c^2} \eta^2 \sin^2\varphi_0 - \gamma_0^{-2}
     \left( \frac{p}{p_0} \right)^{1/2} \right\}_,^{1/2}
\label{eq.4.4}
\end{equation}

\noindent in which $\eta \! \equiv \! r_0/r$. The sign of ${\rm d}\eta
/ {\rm d}\varphi$ in eq.(\ref{eq.4.4}) varies as the jet crosses
the cloud.  For example, for the ultrarelativistic case, in which a
straight trajectory is expected, it is positive for $\eta_* \! > \!  1 /
\sin\varphi_0$, and negative when the inequality is inverted.  A general
analytic solution of eq.(\ref{eq.4.4}) can be found by noting that for
high relativistic velocities  the third term on the right hand side of
eq.(\ref{eq.4.4}) is important only for $\eta \! = \! 1 / \sin\varphi_0$.
In other words, the pressure stratification of the cloud can be written
as eq.(\ref{eq.3.9}) with $\Gamma \!  \rightarrow \!  1/2 $. We can
therefore expand eq.(\ref{eq.4.4}) around $\eta \! = \! 1/ \sin\varphi_0$
to obtain a relation which is the same as eq.(\ref{eq.3.11}) but with:

\begin{equation}
   a = 1 -  \frac{ \alpha }{ \gamma_0^{2} } , \thickspace b = - \frac{
      \beta }{ \gamma_0^{2} } \sin\varphi_0,\thickspace e = -\left( 
      \frac{ \zeta }{ \gamma^2 } + 1 \right) \sin^2
      \negthinspace \varphi_0.                     
\label{eq.4.5}
\end{equation}

\subsection{Isothermal Cloud}
\label{isorel}

   As in section \ref{isoclass}, consider the case in which the jet
propagates through an isothermal cloud In this case it is possible to
find an exact solution to the problem, since eq.(\ref{eq.3.15}) and
eq.(\ref{eq.4.4}) imply:

\begin{equation}
  \frac{{\rm d}\eta}{{\rm d}\varphi} = \pm \frac{1}{ \sin \varphi_0 }
     \left\{ \frac{c^2}{v^2_0} \left( 1-\eta\gamma_0^{-2} \right)
     - \eta^2 \sin^2\negthinspace \varphi_0 \right\}_.^{1/2}
\label{eq.4.6}
\end{equation}

\noindent In other words, the solution is the same as the one already found
in eq.(\ref{eq.3.11}) and eq.(\ref{eq.3.12}) but with:

\begin{equation}
   a=\left(\frac{c}{v_0}\right)^2, \qquad   b=-\left( \frac{ c }{ v_0 }
      \right)^2 \gamma^{-2},  \qquad   e=-\sin^2\varphi_0.
\label{eq.4.7}
\end{equation}

\begin{figure}
   \begin{center}
      \epsfig{file=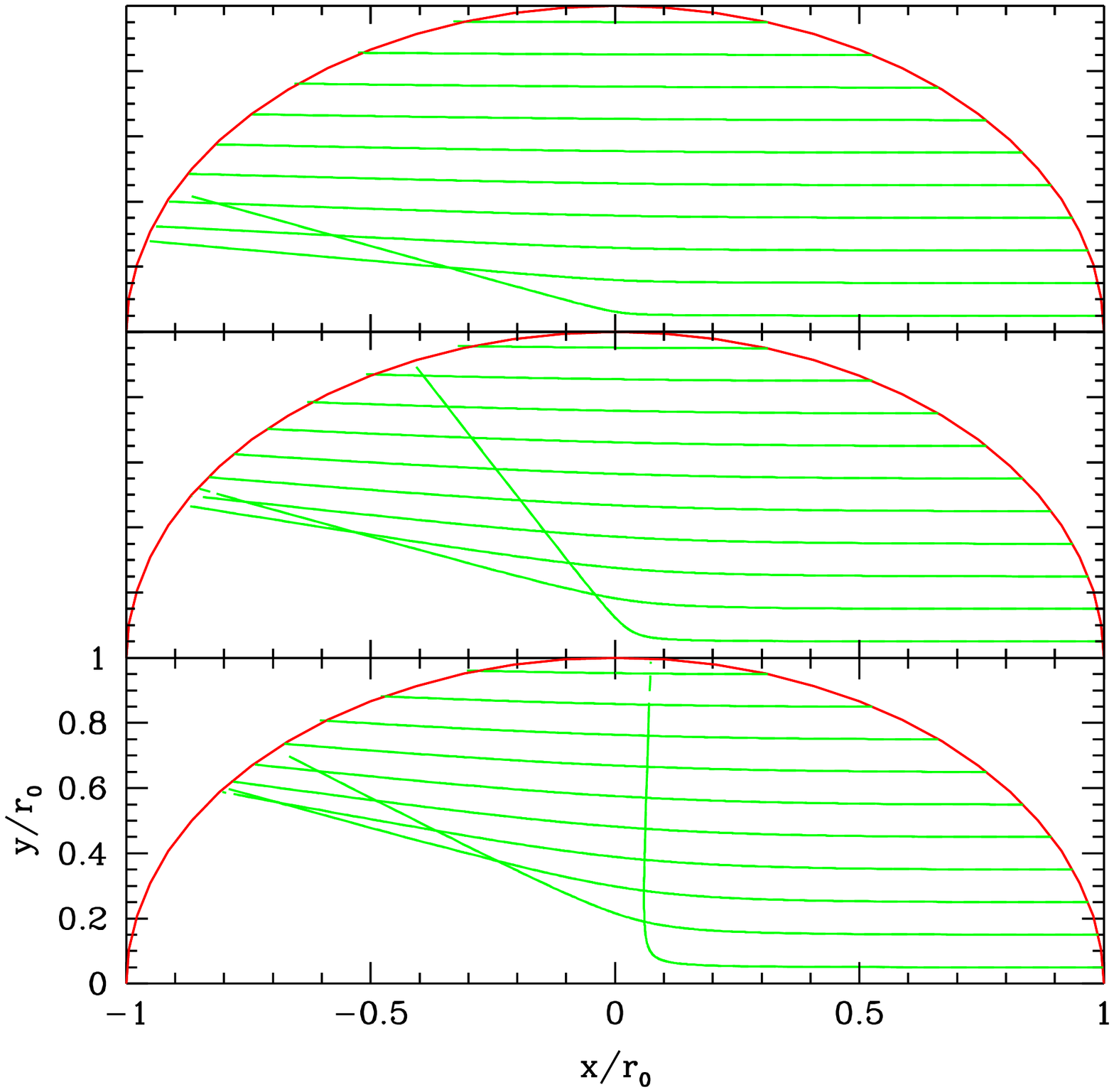,height=8.45cm}
   \end{center}
    \caption{Deflection of a relativistic jet produced by its interaction
	     with an isothermal cloud (semicircle).  The jet is assumed
	     to travel parallel to the $x$ axis at the moment it enters
	     the cloud from the right.  In each plot different
	     trajectories are shown for different values of the initial
	     height of the jet $y/r_0\!=\!0.05$, $0.15$,...,$0.95$ as
	     measured from the $x/r_0$ axis.  The top, central and
	     bottom panel plots were calculated for values of the initial
	     velocity of the jet $v_0$ in units of the speed of light $c$
	     of $0.99$, $0.97$, $0.95$ respectively.}
\label{fig.4.1}
\end{figure}

\noindent Fig.(\ref{fig.4.1}) shows plots of the trajectory of the jet
for  different values of the initial velocity of the jet.

\subsection{Gas on a Dark Matter Halo}
\label{rel-dark-matter}

\begin{figure}
   \begin{center}
      \epsfig{file=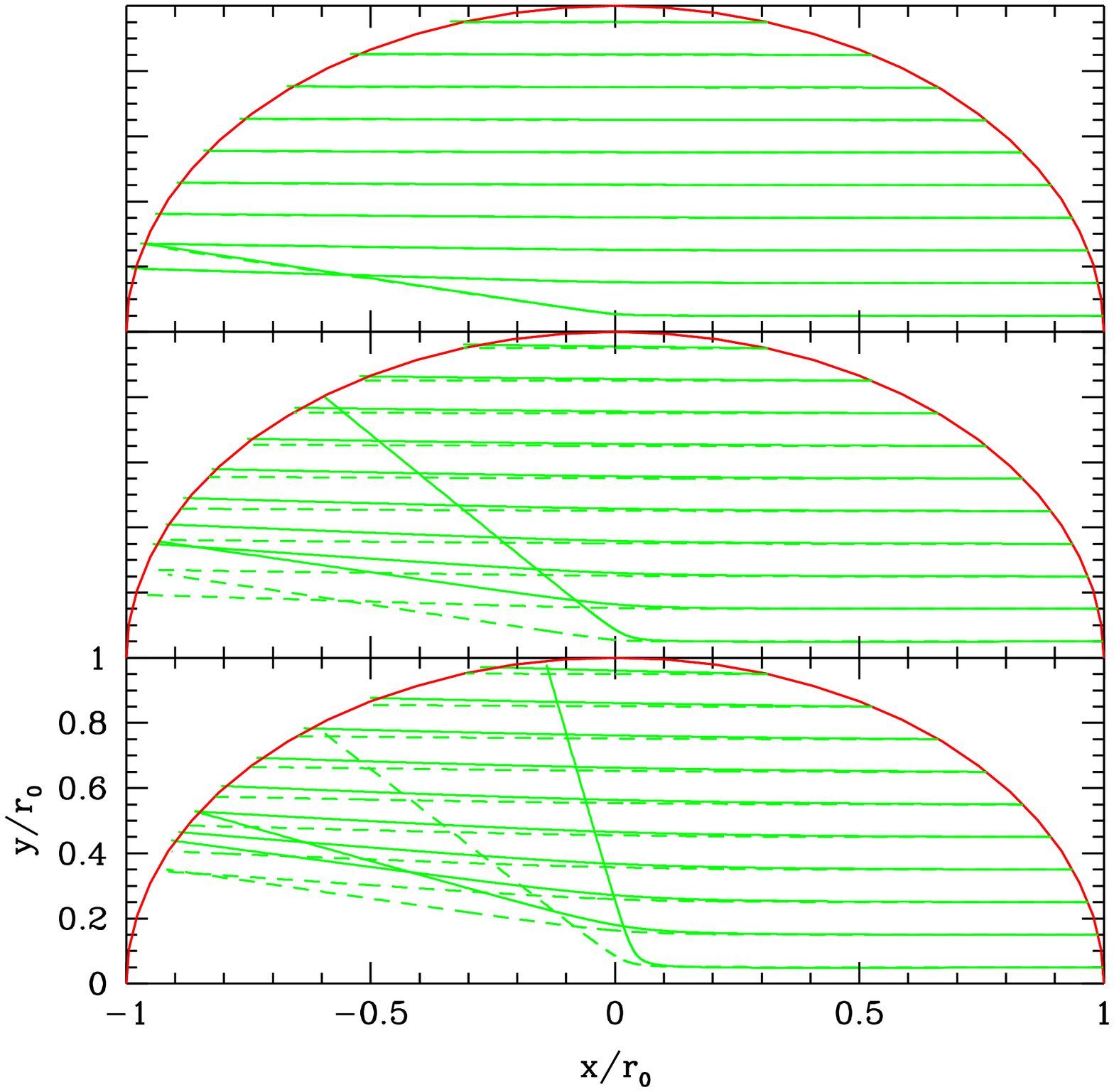,height=8.45cm}
   \end{center}
    \caption{Trajectory of a jet as it crosses a galaxy.  The gas in the
             galaxy is assumed to be in hydrostatic equilibrium with
             a gravitational potential given by a dark matter halo in
             the galaxy.  It is assumed that the jet enters the galaxy
             parallel to the $x$ axis at a height of $y/r_0\!=\!0.05$,
             $0.15$,...,$0.95$ in different cases.  The plots were
             calculated for the case in which the parameter $k \! = \! -3$
             and the initial velocity of the jet in units of the speed
             of light is $0.999$, $0.995$ and $0.99$ from top to bottom.
             Continuous lines are analytic approximations for a high
             Mach number of the flow inside the jet.  The dashed lines
             are direct numerical solutions.}
\label{fig.4.2}
\end{figure}

  As in section \ref{classdark} for $r_0 \! \gg \! a$,  the variation
of the pressure in the galaxy is given by eq.(\ref{eq.3.22}) and the
trajectory of the path of the jet is given by eq.(\ref{eq.3.11}) and
eq.(\ref{eq.3.12}) together with eq.(\ref{eq.3.23}) and the substitution
$\Gamma \! \rightarrow \! 1/2$.  Fig.(\ref{fig.4.2}) shows plots of this
for $k \! = \! -3$ and different values of the initial velocity of the
jet $v_0$.

\section{Discussion}

\begin{figure*}
   \begin{center}
      \epsfig{file=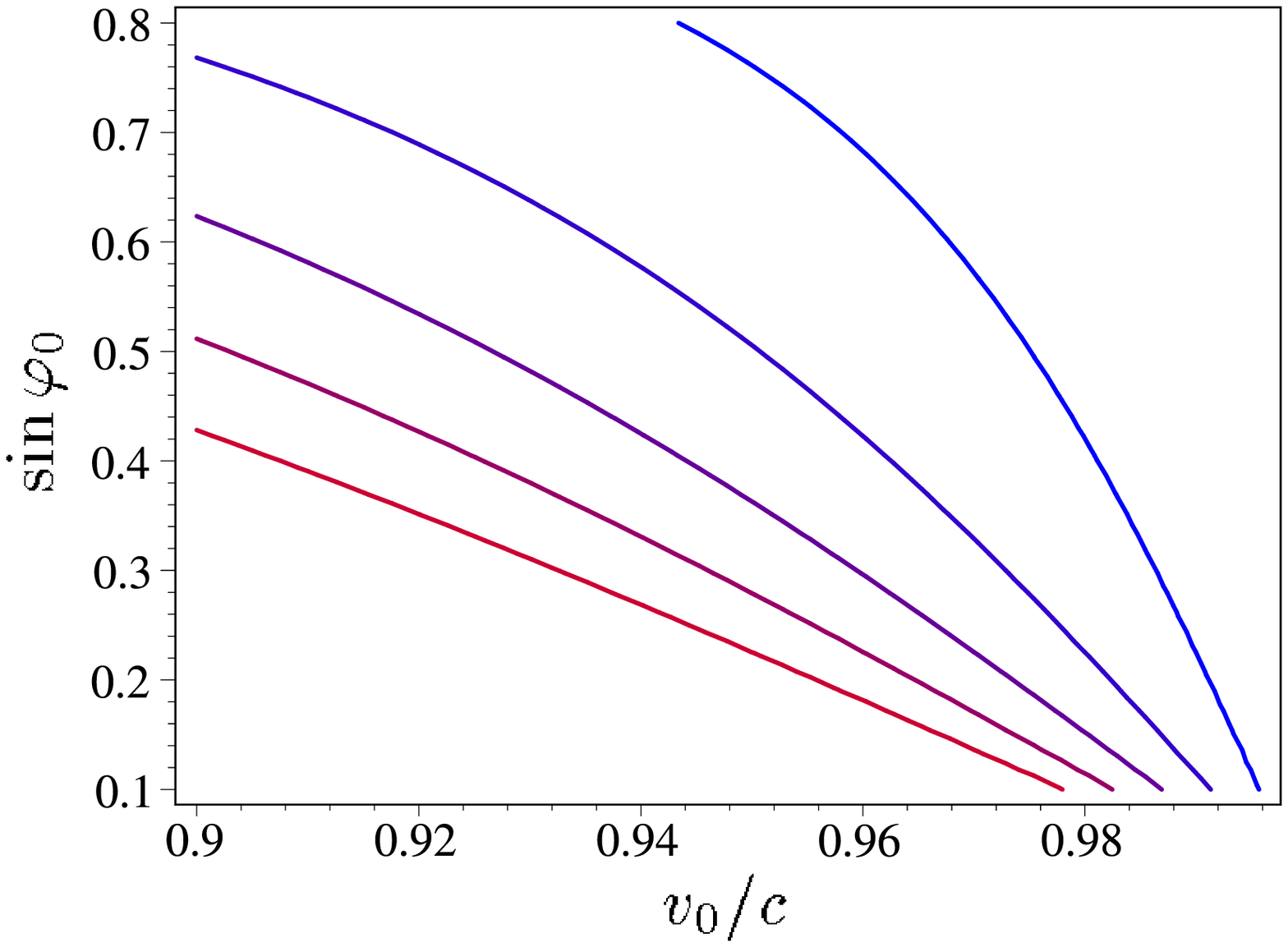,height=6.0cm}
      \epsfig{file=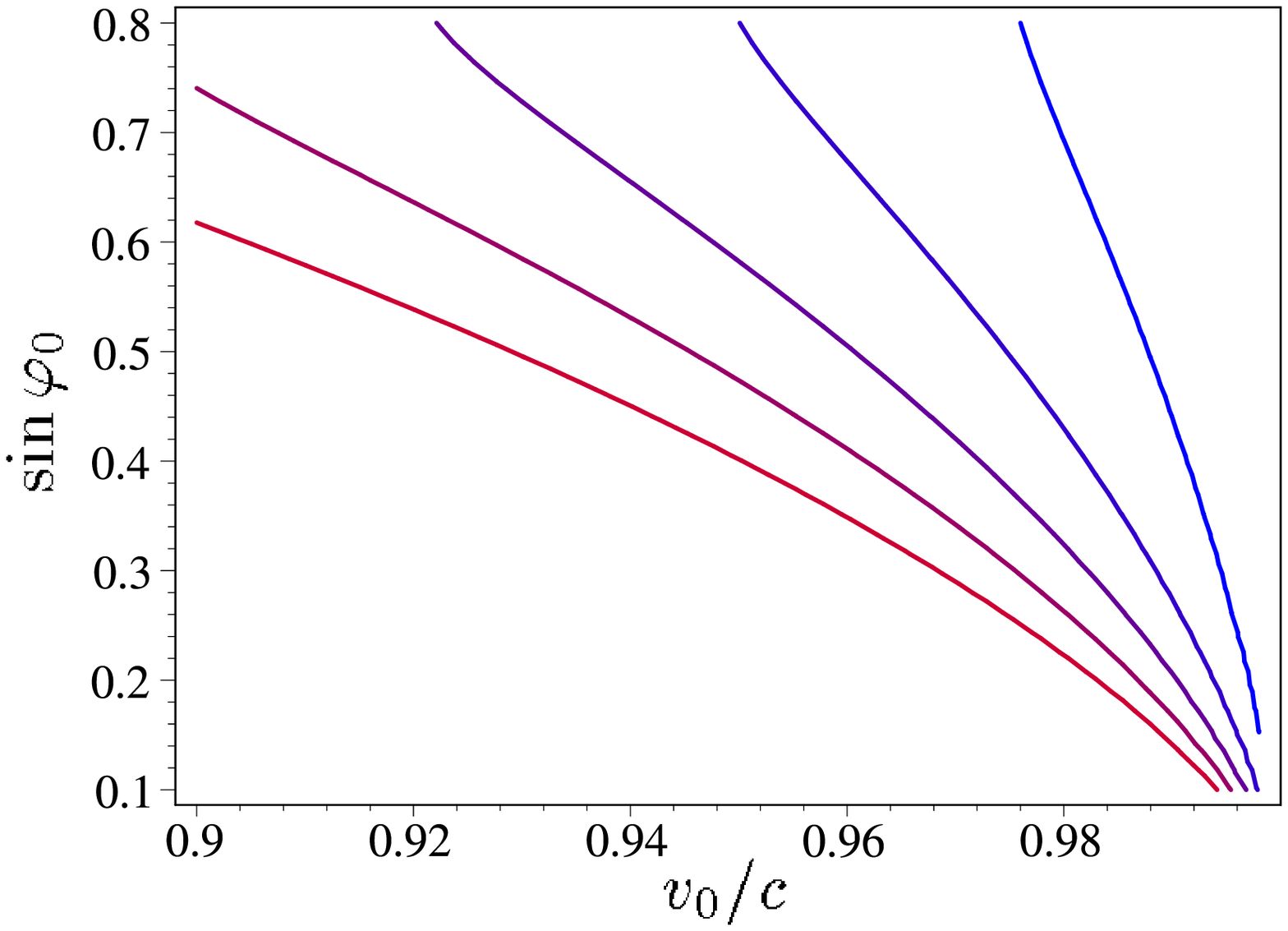,height=6.0cm}
   \end{center}
    \caption{Variations of initial azimuthal angle $\varphi_0$ as a
             function of velocity $v_0$ in units of the speed of light
             $c$ for constant values of the deflection angle $\psi$.
             The angle $\psi$ is the azimuthal angle the velocity
             vector of the flow in the jet makes with the $x$ axis at
             the moment it leaves the cloud.  Every plot was calculated
             for $\cos\psi \! = \! const$ with values given by  $\psi$
             of $ 175^\circ \negthickspace $, $ 170^\circ \negthickspace
             $,...,$ 155^\circ $. The gradient of $\psi$ decreases
             towards the lower left part of each diagram.  In other
             words, deflections become stronger as the curves approach
             this region on the diagram.  An isothermal sphere and
             an isothermal gas in hydrostatic equilibrium with a dark
             matter halo were assumed for the left and right diagrams
             respectively in the case of a  relativistic jet.}
\label{fig.5.1}
\end{figure*}

   When a deflection is observed, it  is possible to work backwards and
find useful properties concerning the initial interaction of a jet with a
stratified density region.  For example, by taking the``standard'' values
used in sections \ref{isoclass} and \ref{classdark} for the pressure and
density in the stratified gas it is possible to calculate the initial
azimuthal angle $\varphi_0$ for a given initial velocity of the jet.
Indeed, from eq.(\ref{eq.3.13}) and eq.(\ref{eq.3.14}), together with
the negative derivative $\left( {\rm d}\eta / {\rm d} \varphi \right)_e
$  at the moment the jet leaves the cloud we can find the value of the
deflection angle $ \cos\psi $ as a function of the velocity of the jet
$v_0$ and the initial azimuthal angle $\varphi_0$.  The curves for
which $ \cos\psi \! = \! \mathrm{const}$ give the required relation
between the initial velocity and azimuthal angle.  Fig.(\ref{fig.5.1})
shows two examples of this type of analysis.

  Different combinations of the various parameters involved (or the
known observables) in  the problem can be formed so that, for a given
deflection, the other quantities can be calculated.  For instance one
can ask for the values of the  central density of the gas in the cloud,
the density in the jet, etc.

  The assumption of pressure equilibrium between the jet and the
external environment has certain limitations on the geometry of the
jet \cite{raga96}.  In the non--relativistic limit pressure equilibrium
is achieved, when the time for a pressure scale height (\( p(r) / |
\mathrm{d} p(r) / \mathrm{d} r | \)) to change significantly within the
jet material is greater than the sound crossing time of the jet radius:

\begin{equation}
  \frac{ \Delta_{\rm j} }{ c_{\rm j} } \ll \frac{ 1 }{ v_{\rm j}
    } \frac{ p(r) }{ | {\rm d} p(r) / {\rm d} r | },
\label{eq.5.1}
\end{equation}

\noindent where \( c_{\rm j} \) is the speed of sound in the jet,
\( v_{\rm j} \) its velocity and \( \Delta_{\rm j} \) its radius
\cite{raga96}.  For the case of an isothermal sphere the right hand side
of eq.(\ref{eq.5.1}) is \( r / 2 v_{ \rm j } \) and for the case of gas
in pressure equilibrium with a dark matter halo is \( r / |k| v_{\rm j} \).
In the case of relativistic flow, eq.(\ref{eq.5.1}) is still valid but
with the substitution \( v_{\rm j} \rightarrow \gamma_{v_{\rm j}} v_{\rm
j} \) and \( c_{\rm j} \rightarrow \gamma_{c_{\rm j}} c_{\rm j} \), where
\( \gamma_{v_{\rm j}} \text{ and } \gamma_{c_{\rm j}} \) are the Lorentz
factors of the flow velocity and the velocity of sound respectively.

  When a jet bends it is in direct contact with its surroundings and
entrainment from the external gas might cause disruption to its structure
\cite{icke91}. However, if this situation is bypassed for example, by an
efficient cooling, then there remains a high Mach number collimated flow
inside a curved jet. When supersonic flow bends, the characteristics
emanating from it intersect at a certain point in space \cite{daufm}.
Since the hydrodynamical values of the flow in a characteristic line
have constant values, the intersection causes the different values of
these quantities to be multivalued.  This situation can not occur in
nature and a shock wave is formed.

  The formation of internal shocks inside the jet gives rise to subsonic
flow inside the jet and collimation may no longer be achieved.  If the
characteristic lines produced by the flow inside the jet intersect outside
the jet, then a shock wave is not formed and the jet remains collimated
as it bends.  However, two important points have to be considered in the
discussion.  The first is that the Mach number decreases in a bend as the
flow moves.  The second is that the rate of change of the Mach angle with
respect to the angle the jet makes with its original straight trajectory
(the bending angle)  increases without bound as the velocity of the flow
tends to that of the local velocity of sound.  This was first proved by
Icke \shortcite{icke91} for the case in which no relativistic effects
were taken into account.  We have made a relativistic generalisation to
these two points which will be discussed in a future paper.  The increase
in the Mach angle means that there has to be a stage at which this angle
grows faster than the bending angle. This causes the characteristics in
the jet to intersect and to form a shock.  The only way to avoid this
situation is if the rate of change of the Mach angle with respect
to the bending angle is less than one.  The formation of this shock
is most probably not too destructive for the jet.  This is because the
Mach number of the flow is close one, which implies that the shock is weak.

  The fact that two shocks might form inside a jet, one of them at the
end of the bending when the Mach number is near one, enables us to find
an upper limit to the bending angle \cite{icke91}.  For example,
a classical jet with $\gamma \! = \! 5/3$ cannot be deflected more than
$60.8^\circ$.  Under the same classical conditions, but by assuming a
polytropic index $\gamma \! = \! 4/3$, classical jets with a relativistic
equation of state cannot be deflected more than $148.12^\circ$
\cite{icke91}.  

  When a full relativistic analysis is introduced in this description,
this upper limit can not be greater than its classical counterpart.
The relativistic Mach angle \cite{konigl80} decreases as the velocity
of the flow approaches that of light.  This means that characteristic
lines emanating from a relativistic flow are closer to the streamlines as
compared with their classical counterparts.  The fact that characteristic
lines are closer to the streamlines means that when a bending occurs, the
chance for an intersection between different characteristics increases.
As a result, a relativistic jet with a polytropic index $ \gamma \! =
\! 4/3 $ can not bend more than $47.94^\circ$.  The precise conditions
under which an internal shock is produced for a given jet depend on the
shape of the curve that the jet makes as it bends and the radius of the
jet. The general conclusion is that, provided the jets are sufficiently
narrow, the deflections shown in figs.(\ref{fig.3.1})-(\ref{fig.4.2})
are valid for small deflections, say \( \lesssim 20\text{--}30^\circ \).

  As an example, we can take the case of the western jet in the
radio galaxy 3C34 \cite{best97}.  If this jet was bent because of
the interaction with a typical galaxy for which its gas is in pressure
equilibrium with a dark matter halo, then we can use the standard values
presented in section \ref{rel-dark-matter}.  First of all, the western
jet in 3C34 has a bending angle \( \Theta \sim 10^\circ \), so that the
deflection angle \( \psi \sim 170^\circ \).  If we assume that the flow
inside the jet moves with a velocity \( v = 0.99 c \), it follows from
the right diagram in fig.(\ref{fig.5.1}) that \( \sin \varphi_0 \sim 0.3
\), or \( \varphi_0 \sim 17^\circ \).  On the other hand, the value \(
\Theta \sim 10^\circ \) is well below the upper limit of \( 47.94^\circ
\) discussed above, so that at least no terminal shock will be produced
by the deflection of the jet.  From preliminary results that will be
published elsewhere, it follows that if the trajectory of the jet in 3C34
is circular, then in order not to produce an internal shock at the onset
of the curvature, the ratio \( D / R \)  (\( D \) is the width of the
jet and \( R \) its radius of curvature) has to be less \( 0.08 \).  In
other words, the destruction of jets only occurs if the jets are thick,
rather than narrow.  This result is in accord with the narrowing of jets
observed in a wide range of different sources.

  The most important consequence of the calculations presented in this
article  is the sensitivity  of the deflection angles to variations
in velocity (see for example fig.(\ref{fig.5.1})).  This sensitivity
is due to the fact that the force applied to a given fluid element
in the jet (due to pressure and gravitational potential gradients) is
the same independent of the velocity of the flow in the jet. However,
as the velocity of the flow in the jet increases, there is not enough
time for this force to change the curvature of the jet soon enough,
giving rise to very straight jets. 

\section{Acknowledgements}

   We would like to thank P. Scheuer for useful comments during the
preparation of this article.  S. Mendoza thanks  support granted by
Direcci\'on General de Asuntos del Personal Acad\'emico (DGAPA) at the
Universidad Nacional Aut\'onoma de M\'exico (UNAM).


\label{lastpage}
\end{document}